\documentclass{article}

\usepackage{authblk}
\usepackage{amssymb}
\usepackage{amsmath}
\usepackage{amsopn}
\usepackage{pifont}
\usepackage{color}
\usepackage{mathrsfs}
\usepackage{amsthm}
\usepackage{graphicx}
\usepackage{hyperref}
\usepackage{color}
\usepackage[numbers,square,sort&compress]{natbib}
\usepackage{pdfpages}

\numberwithin{equation}{section}

\theoremstyle{plain}
\newtheorem{thm}{Theorem}[section]

\newtheorem{cor}[thm]{Corollary}

\theoremstyle{definition}

\usepackage[normalem]{ulem} 
\usepackage{cancel}

\newcommand{\CC}{\mathbb{C}}

\newcommand{\be}[1]{\begin{equation}\label{#1}}
\newcommand{\ba}[1]{\begin{multline}\label{#1}}
\newcommand{\ee}{\end{equation}}
\newcommand{\ea}{\end{eqnarray}}

\newcommand{\tr}{\mathop{\rm tr}}

\newcommand{\Id}{\mathbb{I}}

\newcommand{\bu}{\bar u}
\newcommand{\bv}{\bar v}

\newcommand{\bz}{\bar z}

\begin{document}

\thispagestyle{empty}
\setcounter{page}{1}

\title{Full counting statistics for twisted XXX spin chains}
\author[1]{S. Belliard \thanks{samuel.belliard@univ-tours.fr}}
\affil[1]{Institut Denis Poisson,  CNRS - UMR 7013, Université de Tours, Parc de Grandmont, 37200 Tours, France}
\author[2]{A.Hutsalyuk \thanks{hutsalyuk@gmail.com}}
\affil[2]{SISSA and INFN, Sezione di Trieste, via Bonomea 265, 34136 Trieste, Italy}

\maketitle

\begin{abstract}
Full counting statistics for an arbitrary spin operator is considered for the twisted XXX spin one-half chain. We use the quantum inverse scattering formalism and the modified algebraic Bethe ansatz to construct an explicit formula, given by a form factor expansion.
\end{abstract}

\section{Introduction \label{Intro}}

{\it Full counting statistics} (FCS), pioneered in \cite{LevitovLesovik}, provides access to 
important information about the statistical properties of the system \cite{Schonhamer,IvanovAbanov,CherngDemler,EspositoHarbolaMukamel}. Recently, it was intensively studied in spin chains \cite{BCCKR, GEC, GECol, CColGdGMur, ARV, StephanPollmann,FendleyLamacraft,NR}. We consider FCS for one of the simplest integrable models, the one-dimensional isotropic Heisenberg chain with $L$ spins one-half, given by 
\be{Heisenberg}
H_K=J\sum_{j=1}^L\left(\sigma_j^x\otimes\sigma_{j+1}^x+\sigma_j^y\otimes\sigma_{j+1}^y+\sigma_j^z\otimes\sigma_{j+1}^z\right),
\ee
with $J$ measuring the strength of the exchange interaction and the Pauli matrices $\sigma_j^\alpha$\footnote{
 $
\sigma^{x}=\left(\begin{array}{cc}
       0 & 1\\
      1 & 0      \end{array}\right),\quad
\sigma^{y}=\left(\begin{array}{cc}
       0 & -i\\
      i & 0      \end{array}\right),\quad
\sigma^{z}=\left(\begin{array}{cc}
       1 & 0\\
      0 & -1      \end{array}\right)$.} 
 act non-trivially only in the space $V_j=\CC^2$ of the Hilbert space $\mathcal{H}=\otimes_{j=1}^LV_j$.  
We consider {\it twisted boundary conditions} parametrized by $\{\kappa_1,\kappa_2,\kappa_\pm\}$ and boundary relations :
\be{boundary}
\begin{split}
&\sigma^x_{L+1}=\frac{\kappa_1^2+\kappa_2^2-\kappa_+^2-\kappa_-^2}{2\gamma}\sigma^x_1+\frac{\kappa_2^2-\kappa_1^2+\kappa_+^2-\kappa_-^2}{2i\gamma}\sigma^y_1+\frac{\kappa_2\kappa_--\kappa_1\kappa_+}{\gamma}\sigma^z_1,\\
&\sigma^y_{L+1}=\frac{\kappa_2^2-\kappa_1^2+\kappa_+^2-\kappa_-^2}{2i\gamma}\sigma^x_1+\frac{\kappa_1^2+\kappa_2^2+\kappa_+^2+\kappa_-^2}{2\gamma}\sigma^y_1+\frac{\kappa_2\kappa_-+\kappa_1\kappa_+}{i\gamma}\sigma^z_1,\\
&\sigma^z_{L+1}=\frac{\kappa_2\kappa_+-\kappa_1\kappa_-}{\gamma}\sigma^x_1+i\frac{\kappa_1\kappa_-+\kappa_2\kappa_+}{\gamma}\sigma^y_1+\frac{\kappa_1\kappa_2+\kappa_+\kappa_-}{\gamma}\sigma^z_1,
\end{split}
\ee
here $\gamma=\kappa_1\kappa_2-\kappa_+\kappa_-$.
We consider the operator, 
\be{Q-def}
\exp\left(Q^{(\ell)}(\bar \beta)\right)=\exp\left(\sum_{j=1}^{\ell}Q_{j}(\bar \beta)\right),\qquad \ell\leq L,
\ee
where $Q(\bar \beta)=\left(\beta_x\sigma^x+\beta_y \sigma^y+\beta_z \sigma^z\right)$ and $\bar \beta=\{\beta_x,\beta_y,\beta_z\}$. The FCS can be written as
\be{FCS}
\Big\langle \exp\left(Q^{(\ell)}(\bar \beta)\right)\Big\rangle_{K}=\frac{\langle K|\exp\left(Q^{(\ell)}(\bar \beta)\right)|K \rangle}{\langle K|K\rangle},
\ee
where $|K \rangle$ is a given state (usually the ground state) of $H_K$ with energy $E_K$, {\it i.e.}
\be{}
H_K|K \rangle=E_K|K \rangle.
\ee

The case of $Q_{j}(\bar \beta)=\beta_z \sigma^z_{j}$ in \eqref{Q-def} with a {\it periodic boundary condition $\kappa_i=1$, $\kappa_-=\kappa_+=0$} is already well-studied in the literature \cite{Korepin2,MasterEq,KitanineMailletTerrasSlavnov2008,GohKS04,Kozlowski3,KozlowskiMailletSlavnov5,Q_series,GEC}.
Let us remark that for the XXX spin chain with periodic boundary conditions and for $|K \rangle$ the ground state, due to the $SU(2)$ invariance of the model, the FCS of any $Q_{j}(\bar \beta)$ leads to the same result.    
Cases with {\it twisted boundary conditions} that break the $SU(2)$ invariance have been inaccessible for a long time.

The goal of this paper is to use recent results of the Modified Algebraic Bethe Ansatz (MABA) that provide eigenvalues, Bethe vectors \cite{Belliard-Slavnov-mod}, and the scalar product \cite{BelliardPimenta-overlap,BelliardSlavnov-overlap-0,LinearSB} to express this formula in terms of the Bethe parameters, the solutions of the Bethe equations.

Despite some technical developments required here, the core idea is quite simple: since operator $\sigma^x$ (or $\sigma^y$) breaks $U(1)$ symmetry of the model, it is a natural choice to employ in computations the basis where $U(1)$ symmetry is already broken. Then, by a proper choice of this basis, it is possible to find a case where, say, $\sigma^x$ is a diagonal operator. The same applies for twisted boundary conditions: since such conditions break $U(1)$ symmetry as well, it is quite an intuitive idea to apply a basis with broken $U(1)$ symmetry. The basis with broken $U(1)$ symmetry can be described by the modified Bethe ansatz that we employ below. Such models were also considered by means of the off diagonal Bethe ansatz \cite{WYCS} or the separation of variables method \cite{NPT}.

The paper is organized as follows: in section \ref{QISM}, we introduce the quantum inverse scattering formalism and twisted transfer matrices. In section \ref{MABA}, we recall the results of the MABA to characterize the eigenvalues and eigenvectors of the twisted transfer matrices. In Section \ref{Overlap}, we provide an overlap in terms of a determinant formula. In Section \ref{Corr} we obtain the FCS in integral form. Then, in section \ref{Num}, we provide some numerical results for small chains. Finally, in section \ref{Con}, we discuss some open questions related to our results.   

In the paper, the following functions and notations are used
\be{f-g}
f(u,v)=\frac{u-v+c}{u-v},\qquad g(u,v)=\frac{c}{u-v},\qquad h(v,u)=\frac{f(u,v)}{g(u,v)}, \qquad t(u,v)=\frac{g(u,v)^2}{f(u,v)}.
\ee
The shorthand notation for sets $\bu=\{u_1,\dots,u_a\}$ and $\bu_{k}=\{u_1,\dots,u_a\}\setminus \{u_k\}$ is applied; we note the cardinalities of these sets $|\bu|=a$, $|\bu_k|=a-1$ . For the products of functions or commuting operators over set\footnote{For cases when no product is involved, notations with a vertical bar or a semicolon are used, e.g., $\Lambda(v|\bu)$ or $S_n(\bv;\bu)$. In this case, $\bu$ simply stands for the set of variables on which the function depends.}, we use the notation
\be{short_prod}
B(\bar u)=\prod_{i=1}^{|\bar u|}B(u_i), \qquad f(x,\bar u)=\prod_{i=1}^{|\bar u|}f(x, u_i), \qquad f(\bar v,\bar u)=\prod_{i=1}^{|\bar u|}\prod_{j=1}^{|\bar v|}f(v_j, u_i), \dots
\ee

\section{FCS in the quantum inverse scattering formalism\label{QISM}}

Let us recall the main ingredient of the quantum inverse scattering formalism; an unfamiliar reader can look \cite{BIK,Slavnov-Bethe-review,Slavnov-review} and the references therein.

Integrability relies on the so-called Yang-Baxter equation 
\be{YBE}
R_{ab}(u-v)R_{ac}(u-w)R_{bc}(v-w)=R_{bc}(v-w)R_{ac}(u-w)R_{ab}(u-v).
\ee
All operators act on the space $V_a \otimes V_b\otimes V_c$. Its solutions are called R-matrices and act non-trivially on the space corresponding to their subscripts, {\it e.g.} $R_{ab}$ act on $V_a \otimes V_b$.
For the XXX spin chain, we consider the rational R matrix
\be{R-mat}
R_{ab}(u)=\frac{u}{c}I_{ab}+P_{ab},
\ee
with $P_{ab}=\sum_{i,j}E^{ij}_aE^{ji}_b$ is the permutation operator of the two spaces $V_a$ and $V_b$, and $I=\sum_{i,j}E^{ii}_aE^{jj}_b$ is the unity operator. The matrices $E^{ij}$ are the usual unit matrix with $1$ at the intersection of the line $i$ and the column $j$, and with $0$ elsewhere. For spin one-half, we have $V=\CC^2$, and $i$, $j$ belonging to \{1,2\}.  
\\

This R matrix satisfies the following important properties:
\be{R-prop}
 R_{ab}(0)=P_{ab},\qquad
 R_{ab}(u)^{-1}=\frac{R_{ab}(-u)}{1-u^2},\qquad [K_aK_b,R_{ab}(0)]=0.
\ee
From these properties and the Yang-Baxter equation, it follows that the transfer matrix $t_K(u)$, defined by
\be{tr-K}
t_K(u)=\tr{}_a\Big( K_aR_{a1}(u)R_{a2}(u)\dots R_{aL}(u)\Big),
\ee
forms a family of Abelian quantities
\be{com-t}
[t_K(u),t_K(v)]=0.
\ee
In particular for
\be{twist}
K=
\begin{pmatrix}
\kappa_1& \kappa_+\\
\kappa_-& \kappa_2
\end{pmatrix},
\ee
we can extract the Hamiltonian \ref{Heisenberg} from the transfer matrix using the usual formula:
\be{Hamiltonian}
H_K=c\left.\frac{d}{du}\ln(t_K(u))\right.\Big|_{u=0}+L/2.
\ee
Therefore, the Hamiltonian  and the transfer matrix have the same eigenstates. We denote the eigenvalues of the transfer matrix
\be{tact}
t_K(u)|K \rangle=\Lambda_K(u)|K \rangle
\ee
Moreover, for $u=0$ we have
\be{transfer0}
t_{K}(0)=\tr{}_a\Big( P_{a1}P_{a2}\dots P_{aL}\Big) {K}_1=U{K}_1={K}_LU,
\ee
where $U$ is the shift operator that satisfies
\be{shift}
X_iU=UX_{i+1}, \quad X_{L+1}=X_1 
\ee
for any matrix $X$. Moreover, taking the power $l$ of the transfer matrix and using the shift operator property, we have
\be{}
(t_{K}(0))^l=U^l\prod_{i=1}^lK_i. 
\ee
Introducing another transfer matrix $t_{\tilde K}(u)$ with a twist 
\be{}
\tilde K=K \exp(-Q(\bar \beta)),
\ee
we find  that
\be{inverse}
(t_{\tilde K}(0))^{-l}(t_{K}(0))^{l}=\prod_{i=1}^l\tilde K^{-1}_iK_i=\prod_{i=1}^l\exp(Q_i(\bar \beta)).
\ee
This result is known as the Quantum inverse problem \cite{MailletTerras,NiccoliTerras}.
The twist $\tilde K$ is given explicitly by
\be{}
\tilde K=\cosh r\,K-\frac{\sinh r}{r}\tilde Q(\bar\beta)
\ee
\be{}
\tilde Q(\bar\beta)=
\begin{pmatrix}
\kappa_+\beta_x+i\kappa_+\beta_y+\kappa_1 \beta_z &\kappa_1\beta_x-i\kappa_1\beta_y-\kappa_+ \beta_z\\
\kappa_2\beta_x+i\kappa_2\beta_y+\kappa_- \beta_z&\kappa_-\beta_x-i\kappa_-\beta_y-\kappa_2 \beta_z
\end{pmatrix}
\ee
With $r^2=Q(\bar \beta)^2$. Thus, the operator \eqref{Q-def} can be written in terms of  twisted transfer matrices $t_K(u)$ and $t_{\tilde K}(u)$.

Let's define $|\tilde K \rangle$ as an eigenstate of the transfer matrix $t_{\tilde K}(u)$ with eigenvalue $\Lambda_{\tilde K}(u)$. Using the resolution of the identity on that basis, 
\be{series-KK}
\Id=\sum_{\tilde K} \frac{|\tilde K \rangle \langle \tilde K|}{\langle \tilde K|\tilde K\rangle}
\ee
where the sum is over all the eigenstates of $t_{\tilde K}(u)$.
It follows that 
\be{FCS-1}
\Big\langle \exp\left(Q^{(\ell)}(\bar \beta)\right)\Big\rangle_{K}=\sum_{\tilde K} \left(\frac{\Lambda_{K}(0)}{\Lambda_{\tilde K}(0)}\right)^l\frac{\langle K|\tilde K \rangle\langle \tilde K|K \rangle}{\langle K|K\rangle\langle \tilde K|\tilde K\rangle}.
\ee

\section{Eigen-problem from MABA \label{MABA}}

Now let us recall the needed result for MABA, more details for the derivation can be found in \cite{Belliard-Slavnov-mod}. The {\it monodromy matrix}
\be{monodromy-1}
T_a(u)=R_{a1}(u)\dots R_{aL}(u)=
\begin{pmatrix}
A(u)& B(u)\\
C(u)& D(u)
\end{pmatrix}_a
,
\ee
that is a matrix in an {\it auxiliary space} $V_a= \mathbb C_2$ and  operators $\{A(u),B(u),C(u),D(u)\}$ act on the Hilbert space $\mathcal{H}$ of the model. It satisfies the RTT relation
\be{YB}
R_{ab}(u-v)T_a(u) T_b(v)=T_b(v)T_a(u)R_{ab}(u-v),
\ee
that contains the commutation relations between the operators $\{A(u),B(u),C(u),D(u)\}$. The space of states of $\mathcal{H}$ can be constructed from the highest weight representation of the RTT algebra with the pseudo vacuum $|0\rangle$ and the actions
\be{a-d-def-R}
A(z)|0\rangle=a(z)|0\rangle,\qquad D(z)|0\rangle=d(z)|0\rangle,\quad C(z)|0\rangle=0,\qquad
\ee
and its dual $\langle 0|$, such that $\langle 0|0\rangle=1$ and
\be{a-d-def-L}
\langle 0|A(z)=\langle 0|a(z),\qquad \langle 0|D(z)=\langle 0|d(z),\quad \langle 0|B(z)=0,\qquad
\ee
with
\be{}
a(u)=\Big(\frac{u+c}{c}\Big)^L,\qquad
d(u)=\Big(\frac{u}{c}\Big)^L.
\ee
To implement the MABA and find the eigenvalues and Bethe vector of the twisted transfer matrix \ref{tr-K}, it is needed to decompose the twist matrix in the following way :
\be{}
K=BDA
\ee
with
\be{D-A-B-mat}
A=\sqrt{\mu}
\begin{pmatrix}
1& \frac{\rho_2}{\kappa_-}\\
\frac{\rho_1}{\kappa_+}& 1
\end{pmatrix},\quad
B=\sqrt{\mu}
\begin{pmatrix}
1& \frac{\rho_1}{\kappa_-}\\
\frac{\rho_2}{\kappa_+}& 1
\end{pmatrix},\quad
D=
\begin{pmatrix}
\kappa_1-\rho_1&0\\
0& \kappa_2-\rho_2
\end{pmatrix},\quad \mu=\frac{1}{1-\frac{\rho_1\rho_2}{\kappa_+\kappa_-}},
\ee
and the condition
\be{rho-define}
\rho_1\rho_2-(\kappa_2\rho_1+\kappa_1\rho_2)+\kappa_+\kappa_-=0.
\ee
We remark that this transformation contains a free parameter. That freedom will be crucial to calculate overlaps between states of two different twists in the next section.   
From this transformation, we can introduce the {\it modified transfer matrix }
\be{transfer2}
t_K(u)=\tr KT(u)=\tr\hat D\bar T(u)=(\kappa_1-\rho_1)\bar{A}(u)+(\kappa_2-\rho_2)\bar{D}(u),
\ee
where the modified monodromy matrix $\bar T(u)$ and modified operators $\{\bar A(u),\bar B(u),\bar C(u),\bar D(u)\}$ are given by
\be{T-bar}
\bar{T}(u)=AT(u)B=
\begin{pmatrix}
\bar{A}(u)& \bar{B}(u)\\
\bar{C}(u)&  \bar{D}(u)
\end{pmatrix}.
\ee
They are linear combinations of the original ones. In particular, we have  the {\it modified creation operator} 
\be{B-bar}
\bar{B}(z)=\mu\left(B(z)+\frac{\rho_1}{\kappa_-}A(z)+\frac{\rho_2}{\kappa_-}D(z)+\frac{\rho_1\rho_2}{(\kappa_-)^2}C(z)\right).
\ee
Similarly, we can also define the modified monodromy matrix and operators for the $\tilde K$ twist given by
\be{T-bar-2}
\tilde{T}(u)=\tilde AT(u)\tilde B=
\begin{pmatrix}
\tilde{A}(u)& \tilde{B}(u)\\
\tilde{C}(u)&  \tilde{D}(u)
\end{pmatrix},
\ee
where we used the same decomposition as for the twist $K$ but put tilde parameters. Thus in that case we have the modified creation operator 
\be{B-til}
\begin{split}
\tilde{B}(z)=\tilde\mu\left(B(z)+\frac{\tilde\rho_1}{\tilde\kappa_-}A(z)+\frac{\tilde\rho_2}{\tilde\kappa_-}D(z)+\frac{\tilde\rho_1\tilde\rho_2}{(\tilde\kappa_-)^2}C(z)\right).
\end{split}
\ee

Then the eigen-problem can be stated in the MABA framework by the following Theorem :

{\thm \label{remark1} \cite{Belliard-Slavnov-mod}  
Imposing the constraint
\be{C1-T1}
\rho_1\rho_2-(\kappa_2\rho_1+\kappa_1\rho_2)+\kappa_+\kappa_-=0,
\ee
the modified Bethe vectors (and dual one's) for the transfer matrix $t_K$ \eqref{transfer2} are given by
\be{}
|\bu,K\rangle=\bar B(\bu)|0\rangle,\qquad \langle K, \bu|=\langle 0|\bar C(\bu).
\ee
The action of  the transfer matrix on Bethe vectors and duals are given by
\be{eigenvectors1}
t_K(u)|\bu,K\rangle=\Lambda_K(u|\bu)|\bu,K\rangle,\qquad
\langle K,\bu|t_K(u)=\Lambda_K(u|\bu)\langle K,\bu|,
\ee
with eigenvalue
\be{eigenvalue1}
\Lambda_K(u|\bu)=g(u,\bu)\mathcal Y_K(u|\bu),
\ee
where
\be{Y1}
\mathcal Y_K(u|\bu)=(-1)^L(\kappa_1-\rho_1)a(u)h(\bu,u)+(\kappa_2-\rho_2)d(u)h(u,\bu)+(\rho_1+\rho_2)a(u)d(u)
\ee
and ${\bar u}$ is a solution of the Bethe equations 
\be{BE1}
\mathcal Y_K(u_j,\bu_j)=0,\quad j=1,\dots,L.\\
\ee
}

{
\cor\cite{Belliard-Slavnov-mod}
Imposing the constraint
\be{C2-T}
\tilde\rho_1\tilde\rho_2-(\tilde\kappa_2\tilde\rho_1+\tilde\kappa_1\tilde\rho_2)+\tilde\kappa_+\tilde\kappa_-=0,
\ee
the modified Bethe vectors (and dual ones) for the transfer matrix $t_{\tilde K}$ \eqref{transfer2} are given by
\be{}
|\bv,\tilde K\rangle=\tilde B(\bv)|0\rangle,\qquad \langle \tilde K, \bv|=\langle 0|\tilde C(\bv).
\ee
The action of  the transfer matrix on Bethe vectors and dual ones are given by
\be{eigenvectors2}
t_K(v)|\bv,{\tilde K}\rangle=\Lambda_{\tilde K}(v|\bv)|\bv,{\tilde K}\rangle,\qquad
\langle {\tilde K} ,\bv|t_{\tilde K}(v)=\Lambda_{\tilde K}(v|\bv)\langle {\tilde K},\bv|,
\ee
with eigenvalue
\be{eigenvalue2}
\Lambda_{\tilde K}(v|\bv)=g(z,\bv)\mathcal Y_{\tilde K}(v|\bv),
\ee
where
\be{Y2}
\mathcal Y_{\tilde K}(v|\bv)=(-1)^L(\tilde\kappa_1-\tilde\rho_1)a(v)h(\bv,v)+(\tilde\kappa_2-\tilde\rho_2)d(v)h(v,\bv)+(\tilde\rho_1+\tilde\rho_2)a(v)d(v)
\ee
and ${\bar v}$ is a solution of the Bethe equations 
\be{BE2}
\mathcal Y_{\tilde K}(v_j|\bv_j)=0,\qquad j=1,\dots,L.\\
\ee
}

\section{Overlap from modified algebraic Bethe ansatz \label{Overlap}}

To calculate the overlap for different twists, we will use the result about {\it Modified Slavnov determinant} \cite{BelliardPimenta-overlap,BelliardSlavnov-overlap-0,LinearSB}, which  provides a determinant formula for the scalar product of an {\it On-shell Bethe vector} and an {\it Off-shell Bethe vector}\footnote{{\it On-shell Bethe vectors} depend on parameters that are solutions of the Bethe equations, and {\it Off-shell Bethe vectors} depend on free parameters.} for the same twist.

The fact that the MABA has a free parameter for each twist allows us to equate {\it off shell} Bethe vectors for different twists,
\be{}
|\bu,K\rangle/\mu^L=|\bu,\tilde K\rangle/\tilde\mu^L.
\ee
Here, we identify modified creation operators $\bar B(u)$ and $\tilde B(u)$ such that $\bar B(u)/\mu=\tilde B(u)/\tilde\mu$, which brings the constraints :
\be{}
\frac{\rho_1}{\kappa_-}=\frac{\tilde\rho_1}{\tilde\kappa_-},\qquad  \frac{\rho_2}{\kappa_-}=\frac{\tilde\rho_2}{\tilde\kappa_-},
\ee
in addition to the MABA one's
\be{}
\rho_1\rho_2+\kappa_-\kappa_+-(\kappa_1\rho_2+\kappa_2\rho_1)=0,\qquad  \tilde\rho_1\tilde\rho_2+\tilde\kappa_-\tilde\kappa_+-(\tilde\kappa_1\tilde\rho_2+\tilde\kappa_2\tilde\rho_1)=0.
\ee

Therefore, the overlaps are given by
\be{scalp-K1K2}
\langle \tilde K, \bv|\bu, K \rangle=\langle 0|\tilde C(\bv)\bar B(\bu)|0\rangle=(\mu/\tilde\mu)^L\langle 0|\tilde C(\bv)\tilde B(\bu)|0\rangle=(\mu/\tilde\mu)^LS_{\tilde K}(\bv|\bu).
\ee
\be{scalp-K1K2-2}
\langle  K, \bu|\bv, \tilde  K \rangle=\langle 0|\bar C(\bu)\tilde B(\bv)|0\rangle=(\tilde\mu/\mu)^L\langle 0|\bar C(\bu)\bar B(\bv)|0\rangle=(\tilde\mu/\mu)^L S_{K}(\bu|\bv).
\ee
Where we can consider both sets ${\bar u}$ and ${\bar v}$ {\it Off shell}.
\\

Then the scalar product can be computed in the first case \ref{scalp-K1K2} putting {\it On shell} the set $\bv$ from the Bethe equations \ref{BE2} for the twist $\tilde K$  and in the second case \ref{scalp-K1K2-2} by putting {\it On shell} the set $\bu$ from the Bethe equations \ref{BE1} for the twist $K$. Then it follows that the two overlaps are given by  Slavnov's determinants:
\be{Mod-mod-overlap-1}
S_K(\bu;\bv)=\langle 0|\bar B(\bu)|0\rangle \frac{\Delta(\bv)\Delta'(\bu)d(\bu)}{ g(\bv,\bu)}\det\left(c\frac{\partial\Lambda_K(v_k|\bu)}{\partial u_j}\right),
\ee
\be{Mod-mod-overlap-2}
S_{\tilde K}(\bv;\bu)=\langle 0|\tilde B(\bv)|0\rangle \frac{\Delta(\bv)\Delta'(\bu)d(\bv)}{ g(\bu,\bv)}\det\left(c\frac{\partial\Lambda_{\tilde K}(u_k|\bv)}{\partial v_j}\right),
\ee
with the Jacobian of the eigenvalue is 
\be{Lambda-Jacobian}
\begin{split}
c\frac{\partial\Lambda_K(v_k|\bu)}{\partial u_j}&=\left[(\kappa_1-\rho_1)a(v_k)t(u_j,v_k)f(\bu,v_k)\frac{}{}\right.\\
&\left.\frac{}{}+(\kappa_2-\rho_2)d(v_k)t(v_k,u_j)f(v_k,\bu)+(\rho_1+\rho_2)a(v_k)d(v_k)g(v_k,\bu)g(v_k,u_j)\right],
\end{split}
\ee
and the skew-symmetric products $\Delta(\bu)$, $\Delta'(\bu)$  are defined in the following way
\be{Delta-def}
\Delta(\bu)=\prod_{i>j}g(u_i,u_j),\qquad \Delta'(\bu)=\prod_{i<j}g(u_i,u_j).
\ee

We will also need the norm of the eigenvectors given by
\be{Norm-mod-K}
N_K(\bu;\bu)=\langle 0|\bar B(\bu)|0\rangle\Delta(\bu)\Delta'(\bu)d(\bu)\det_L\left(c\frac{\partial\mathcal Y_K(u_k|\bu)}{\partial u_j}\right),
\ee
with $\bu$ solution of \ref{BE1} and
\be{Norm-mod-Kt}
N_{\tilde K}(\bv;\bv)=\langle 0|\tilde B(\bv)|0\rangle \Delta(\bv)\Delta'(\bv)d(\bv)\det_L\left(c\frac{\partial\mathcal Y_{\tilde K}(v_k|\bv)}{\partial v_j}\right),
\ee
with $\bv$ solution of \ref{BE2}.

\section{FSC formula \label{Corr}}

We can now rewrite the FCS in the MABA framework. The FCS \ref{FCS-1} can be rewritten as
\be{ff_sum}
\Big\langle \exp\left(Q^{(\ell)}(\bar \beta)\right)\Big\rangle_{K}=\sum_{\bv}\left(\frac{\Lambda_K(0|\bu)}{\Lambda_{\tilde K}(0|\bv)}\right)^{\ell}\frac{S_{K}(\bu|\bv)}{N_K (\bu)}\frac{S_{\tilde K}(\bv|\bu)}{N_{\tilde K} (\bv)}.
\ee
Then, using \ref{Mod-mod-overlap-1},\ref{Mod-mod-overlap-2},\ref{Norm-mod-K}, and \ref{Norm-mod-Kt}, it follows our main result
\be{ff_sum_2}
\Big\langle \exp\left(Q^{(\ell)}(\bar \beta)\right)\Big\rangle_{K}=\sum_{\bv}\frac{1}{g(\bu,\bv)g(\bv,\bu)}\left(\frac{\Lambda_K(0|\bu)}{\Lambda_{\tilde K}(0|\bv)}\right)^{\ell}\frac{\det_L\left(c\frac{\partial\Lambda_K(v_k|\bu)}{\partial u_j}\right)\det_L\left(c\frac{\partial\Lambda_{\tilde K}(u_k|\bv)}{\partial v_j}\right)}{\det_L\left(c\frac{\partial\mathcal Y_K(u_k|\bu)}{\partial u_j}\right)\det_L\left(c\frac{\partial\mathcal Y_{\tilde K}(v_k|\bv)}{\partial v_j}\right)},
\ee
where the sum is taken over all {\it admissible solutions}\footnote{There also exist 
{\it spurious solutions} of the Bethe equations, where each term in \eqref{Y2} is equal to zero or a few rapidities coincide; see details \cite{Slavnov-review}. It was shown that the contributions of such solutions to the sum \eqref{ff_sum_2} are zero \cite{Q_series}.
} $\bar v$ of the Bethe equation \ref{BE2}. 

Above, it was assumed that the Bethe equation \eqref{BE1} parametrizes a complete set of states. We lack the proof of completeness in the case of an arbitrary non-diagonal twist $K$, but it is worth mentioning that completeness has been probed in a similar case of open boundary conditions \cite{Nepomechi03,Nepomechi13}. In general, the proof of the completeness of the Bethe ansatz is a complicated question, addressed even in the case of ordinary ABA in a limited number of cases \cite{MVT,Chernyak}.

We can rewrite the sum as a contour integral using the following trick \cite{Slavnov-Bethe-review}. For an arbitrary function $F(\bar\mu)$, the sum over solutions of equations $\mathcal Y_{\tilde K}(v_j|\bv)=0$ can be rewritten as a contour integral around the solutions of the Bethe equations 
\be{sum_int}
\sum_{\bar v}F(\bar v)=\frac{1}{L!}\oint\limits_{\bar v}\frac{d\bar z}{(2\pi)^L}\:\det\left(\frac{\partial\mathcal Y_{\tilde K}(z_j|\bz)}{\partial z_k}\right)\frac{ F(\bar z)}{\prod\limits_{j=1}^n\mathcal Y_{\tilde K}(z_j|\bz)}.
\ee
Factorial $n!$ appears to avoid multiple counting of the Bethe states that differ only by permutations of the spectral parameters within the set. Thus, we can present the FCS in the form
\be{ff_integral-2}
\Big\langle \exp\left(Q^{(\ell)}(\bar \beta)\right)\Big\rangle_{K}=\frac{1}{L!}\oint\limits_{\bv}\frac{d\bar z}{(2\pi)^L }\frac{1}{g(\bu,\bz)g(\bz,\bu)}\left(\frac{\Lambda_K(0|\bu)}{\Lambda_{\tilde K}(0|\bz)}\right)^{\ell}\frac{\det_L\left(c\frac{\partial\Lambda_K(z_k|\bu)}{\partial u_j}\right)\det_L\left(c\frac{\partial\Lambda_{\tilde K}(u_k|\bz)}{\partial z_j}\right)}{\det_L\left(c\frac{\partial\mathcal Y_K(u_k|\bu)}{\partial u_j}\right)\prod\limits_{j=1}^L\mathcal Y_{\tilde K}(z_j|\bz)}.
\ee

 Representations \eqref{ff_sum} and \eqref{ff_integral-2} are the main results of this paper, with the former being a convenient representation for both numerical and analytic analyses of the long-range behaviour of the FCS \cite{KitanineMailletTerrasKozlowski,KozlowskiMailletSlavnov5,Kozlowski3}, while the latter is suitable rather for short-range computations \cite{GohKS04,Q_series}.

\section{Some numeric \label{Num}}

We present some numerical results for a few cases.
\begin{itemize}
\item {\bf Case 1.} $K=\mathbb \sigma^x$ and expectation values of $\exp(Q^{(\ell)}(\bar\beta))$ with $\bar\beta=\{1,0,1\}$ and $\bar\beta=\{1,1,1\}$ are shown.

\begin{figure}[h]
    \centering
    \includegraphics[width=0.65\linewidth, page=1]{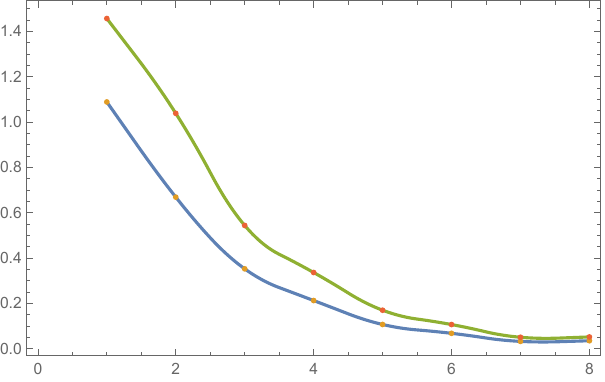}
    \caption{FCS, $K=\sigma^x$, expectation values of $\exp(Q^{(\ell)}(\bar\beta))$ with $\bar\beta=\{1,0,1\}$ (the blue curve) and $\bar\beta=\{1,1,1\}$ (the green curve) are shown, $L=8$. Dots correspond to direct diagonalization, while curves are interpolation of results derived via MABA.}
\end{figure}

\item  {\bf Case 2}. $K=\sigma^y$ and expectation values of $\bar\beta=\{1,0,1\}$, $\bar\beta=\{1,1,1\}$, and $\bar\beta=\{1,-1,2\}$.

\begin{figure}[h!]
    \centering
    \includegraphics[width=0.65\linewidth, page=1]{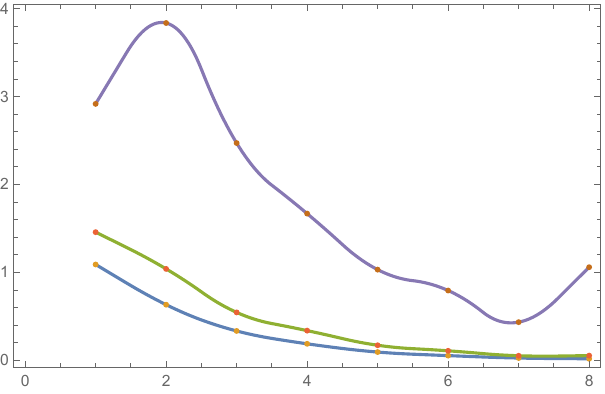}
    \caption{FCS,  $K=\sigma^y$, expectation values of $\exp(Q^{(\ell)}(\bar\beta))$ with $\bar\beta=\{1,0,1\}$ (the blue curve) and $\bar\beta=\{1,1,1\}$ (the green curve) and $\beta=\{1,-1,2\}$ (purple curve) are shown, $L=8$. Dots correspond to direct diagonalization, while curves are interpolation of results derived via MABA.}
\end{figure}

\end{itemize}

We use @Mathematica for direct calculation and for the MABA solution with $L=8$. 
In order to perform the numerical summation of the series \eqref{ff_sum}, all admissible solutions of the Bethe equations should be found. It can be done either by  directly solving the system \eqref{Y1} or by solving an equivalent {\it TQ-equation}
\be{TQ}
t(u)Q(u)=(\kappa_1-\rho_1)a(u)Q(u+c)+(\kappa_2-\rho_2)d(u)Q(u-c)+(\rho_1+\rho_2)a(u)d(u),
\ee
with the transfer matrix and $Q$-polynomial given by
\be{}
t(u)=\sum_{j=0}^L t_ju^j,\qquad Q(u)=u^L+\sum_{j=0}^{L-1}q_ju^j,
\ee
with zeros of $Q(u)$ defining the set of Bethe roots. This is known in the literature as McCoy {\it et. al.} method \cite{FabriciusMccoy}.

\section{Conclusion \label{Con}}

Using new developments in the algebraic Bethe ansatz, we were able to compute the FCS of the spin-1/2 isotropic Heisenberg chain in terms of the form factor sum. We expect this representation to be convenient for both numerical and analytic studies of the long-range asymptotic behavior of the FCS, since methods for the analysis of such series are well developed. In particular, it was performed for a case of $\sigma^z$ operators \cite{Kozlowski3,KitanineMailletTerrasKozlowski,KozlowskiMailletSlavnov5}. In \cite{KozlowskiMailletSlavnov4}, the form factors summation was generalized to the dynamical case; further, finite temperature states were studied \cite{Gohmann1,Gohmann3} as well, using the quantum transfer matrix concept.

The form factor summation in the case of MABA will require modifications to existing methods. In particular, the thermodynamic limit of a single form factor should be taken, similarly to \cite{KitanineMaiilletTerrasSlavnov2009,KitanineMaiilletTerrasSlavnov2010}. Even this task is non-trivial due to the absence of a clear understanding of  the distribution of Bethe rapidities in the thermodynamic limit. Moreover, the summation of the form factor series becomes even more challenging due to the former fact. We plan to approach this problem in the future.

Except for the FCS, the method that is, in fact, nothing but building a form factor series using a convenient basis of twisted Hamiltonians, can be applied to the computation of other quantities. The most obvious ones will be simply products of a type $\sigma_1^{\pm}\dots\sigma_j^{\pm}$; however, other choices, of course, are possible. Similarly to the case of twisted boundary conditions, the choice of open integrable boundary conditions for the state $\tilde K$ in \eqref{series-KK}-\eqref{FCS-1} is possible if the overlaps of their eigenvectors with the eigenstates of the initial states, i.e., $\langle K|\tilde K\rangle$, are known.

One more challenging problem concerns the generalization of the presented approach to different models. While it is rather straightforward to generalize form factor series to the case of integrable models with higher spins related to the rational six-vertex R-matrix, it could be quite a difficult task to generalize it for the case of the anisotropic Heisenberg chain (XXZ model), which is related to the trigonometric R-matrix. The problem is that, in the last case, the commutation relation in \eqref{R-prop} will place restrictions on possible twists $K$, limiting our options to choose a convenient basis.

Finally, it will be an even more interesting task to generalize the existing approach for other integrable models, such as the Hubbard model or, say, models with $SU(N)$, $N>2$ symmetry.

\section*{Acknowledgements}

A.H. would like to thank Filiberto Ares, Colin Rylands, and Fedor Levkovich-Maslyuk for fruitful discussions. A.H. acknowledges the grants PNRR MUR Project PE0000023-NQSTI and PRO3 Quantum Pathfinder.

\end{document}